\documentclass[fleqn,10pt]{wlscirep}
\title{conLSH: Context based Locality Sensitive Hashing for Mapping of noisy SMRT Reads}

\author[1,*]{Angana Chakraborty}
\author[2]{Sanghamitra Bandyopadhyay}

\affil[1]{Department of Computer Science, Sister Nibedita Govt. General Degree College for Girls, Kolkata, India}
\affil[2]{Machine Intelligence Unit, Indian Statistical Institute, Kolkata, India}
\affil[*]{angana\_r@isical.ac.in}

\begin{abstract}
Single Molecule Real-Time (SMRT) sequencing is a recent advancement of Next Gen technology developed by Pacific Bio (PacBio). It comes with an explosion of long and noisy reads demanding cutting edge research to get most out of it. To deal with the high error probability of SMRT data, a novel \emph{con}textual Locality Sensitive Hashing (conLSH) based algorithm is proposed in this article, which can effectively align the noisy SMRT reads to the reference genome. Here, sequences are hashed together based not only on their closeness, but also on similarity of context. The algorithm has $\mathcal{O}(n^{\rho+1})$ space requirement, where $n$ is the number of sequences in the corpus and $\rho$ is a constant. The indexing time and querying time are bounded  by $\mathcal{O}( \frac{n^{\rho+1} \cdot \ln n}{\ln \frac{1}{P_2}})$ and $\mathcal{O}(n^\rho)$ respectively, where $P_2 > 0$, is a probability value.  This algorithm is particularly useful for retrieving similar sequences, a widely used task in biology. The proposed conLSH based aligner is compared with rHAT, popularly used for aligning SMRT reads, and is found to comprehensively beat it in speed as well as in memory requirements. In particular, it takes approximately $24.2\%$ less processing time, while saving about $70.3\%$ in peak memory requirement for \emph{H.sapiens} PacBio dataset.

\end{abstract}

\usepackage{amsthm}
\usepackage{multirow}
\usepackage{float}
\usepackage{amsmath}
\usepackage{amssymb}
\usepackage{algorithm}
\usepackage{algorithmic}
\usepackage{caption}
\usepackage{multicol}
\usepackage{subcaption}
\begin{document}

\flushbottom
\maketitle
\thispagestyle{empty}

\section*{Introduction}
\begin{figure}[ht]
\centering
\includegraphics[scale=.5,trim=4 4 4 4,clip]{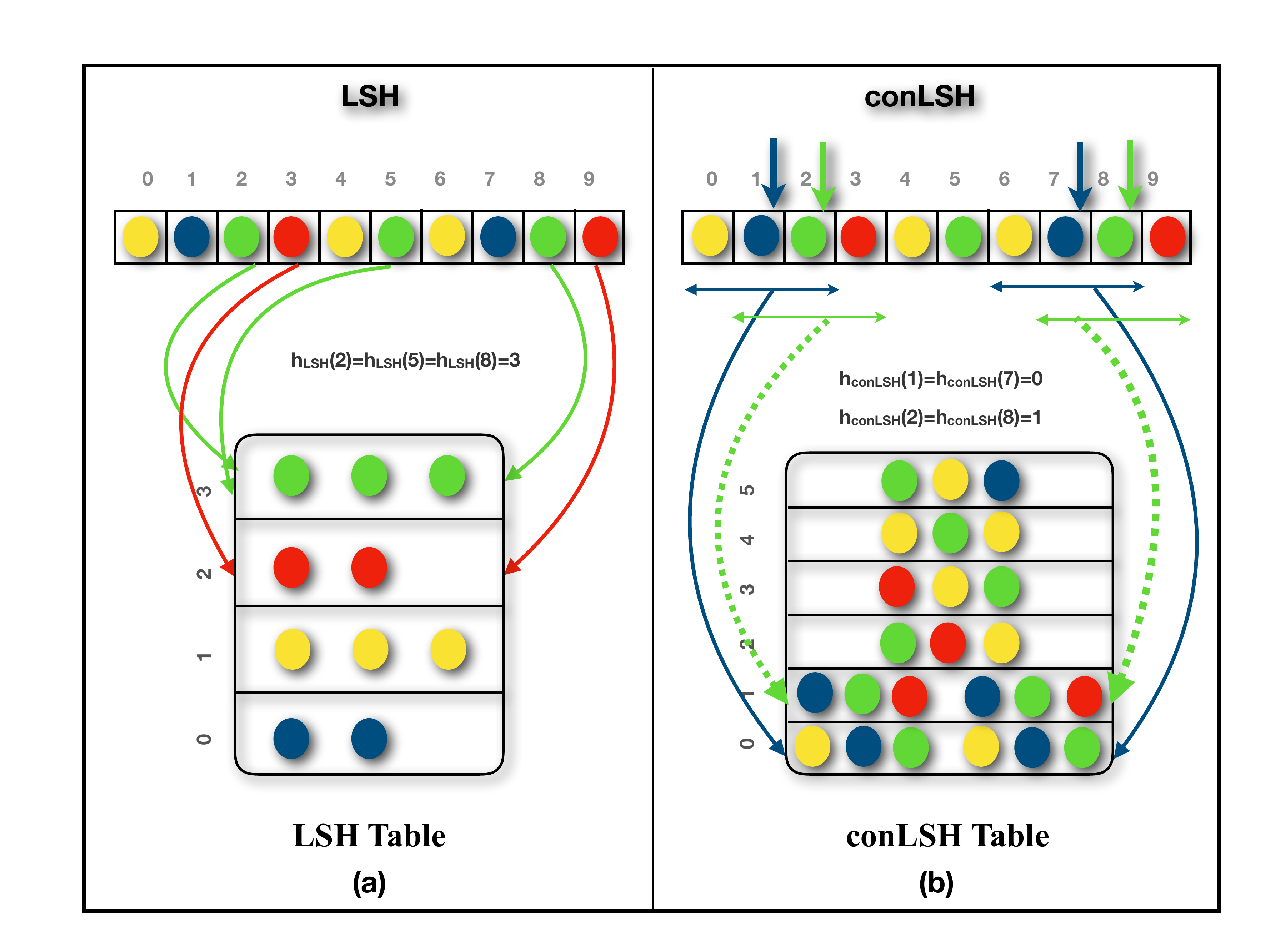}
\caption{Portrays the conceptual difference between Locality Sensitive Hashing and Context based Locality Sensitive Hashing. Figure 1(a) shows that LSH places the balls of same color into the same slots of the hash table. The green balls from positions 2, 5, and 8 are hashed into the 3rd slot of LSH Table. conLSH, on the other hand, groups the green balls from positions 2 and 8 together (Figure 1(b)), keeping the ball from 5 at a different location in the hash table. The reason is that the green balls at positions 2 and 8 share a common context of blue and red in both sides of them. Whereas, the ball from position 5 has a different context of two yellow balls in its left and right, which makes it hashed differently. }
\label{LSH vs conLSH}
\end{figure}

Locality Sensitive Hashing based algorithms are widely used for finding approximate nearest neighbors with constant error probability. The principle of locality sensitive hashing \cite{LSH_old98} originated from the idea to hash similar objects into the same or localized slots of the hash table. That is, the probability of collision of a pair of objects is, ideally, proportional to their similarity. The only things that need to be taken care of are the proper selection of hash functions and parameter values. Several hash functions have already been developed, like those for d-dimensional Euclidean space \cite{LSH_new}, p-stable distribution \cite{LSH_p-stable}, $\chi^2$ distance \cite{LSH_chi2}, angular similarity \cite{jibatch}, etc. Specialized versions of hashing schemes have been applied efficiently in image search \cite{LSH_application3,LSH_application5}, multimedia analysis \cite{LSH_application2,LSH_3D}, web clustering \cite{chakraborty2014layered,LSH_webStructure}, and active learning \cite{LSH_active}, etc. LSH based methods are not new to sequential data. They have profound applications in biological sequences, from large scale genomic comparisons \cite{buhler2001efficient,lshBio_angana} to high throughput genome assembly of SMRT reads \cite{berlin2015assembling}. However, there are certain scenarios, as in sequential data, where the proximity of a pair of points cannot be captured without considering their surroundings or context. LSH has no mechanism to handle the contexts of the data points. In this article, a novel algorithm named Context based Locality Sensitive Hashing (conLSH) has been proposed to group similar sequences taking into account the contexts in the feature space. Here, the hash value of the $i$th point in the sequence is computed based on the values of the $(i-1),i$ and $(i+1)$th strips, when context length is 3. This feature is best illustrated in \ref{LSH vs conLSH}.\\

Single Molecule Real-Time sequencing technology, developed by PacBio using zero-mode waveguides (ZMW) \cite{rhoads2015pacbio}, is able to produce longer reads of length 20KB on  average \cite{ardui2018single}. The increased read length is particularly useful for alignment of repetitive regions during genome assembly \cite{ye2016dbg2olc,khiste2017hisea}. Moreover, SMRT reads possess an inherent capability of detecting DNA methylation and kinetic variations by light pulses \cite{rhoads2015pacbio}. The lowest GC bias of SMRT reads makes them a popular choice for many downstream analysis \cite{roberts2013advantages,ye2016sparc}. However, all these advantages are bundled with one major concern that SMRT reads come with a higher error probability of $13\%-15\%$ per base \cite{ardui2018single}. Unlike second generation of NGS technology, the errors here are mostly indels than mismatches \cite{ardui2018single}. This is the reason why the state of the art NGS read aligners \cite{li2009fast,langmead2012fast,langmead2009ultrafast} are not performing well with SMRT data. The speed of SMRT alignment is, in general, slower than traditional NGS reads \cite{chaisson2012mapping}. The aligners designated for SMRT reads are BWA-SW \cite{li2010fast}, BLASR \cite{chaisson2012mapping} and BWA-MEM \cite{li2013aligning}. All these three methods are based on seed and extension strategy using Burrows Wheel Transformation (BWT). BWA-SW performs suffix trie traversing where BWA-MEM uses exact short token matching for aligning reads to reference genome. BLASR designed a probability based error optimization model to find the alignment. However, these aligners suffer from huge seeding cost to ensure a desired level of sensitivity\cite{liu2015rhat}. An effort has been made to overcome this bottleneck using regional Hash Table (rHAT \cite{liu2015rhat}). The reference window(s) having most kmer matches, as obtained from rHAT index, are considered for extension using Sparse Dynamic Programming (SDP) based alignment. However, rHAT has huge memory footprint (13.70 Gigabytes for \emph{H. Sapiens} dataset \cite{liu2015rhat}) and produces large index file even with the default parameter setting of k=13. Long SMRT reads facilitate the use of longer seeds for sensitive alignment and repeat resolution \cite{chaisson2012mapping}. However, the excessive memory requirement restricts rHAT from increasing kmer size. Herein, we propose a novel concept of Context based Locality Sensitive Hashing (conLSH) to  align noisy PacBio data effectively within limited memory footprint.\\ 

\begin{figure}[ht]
\centering
\includegraphics[scale=.5,trim=4 4 4 4,clip]{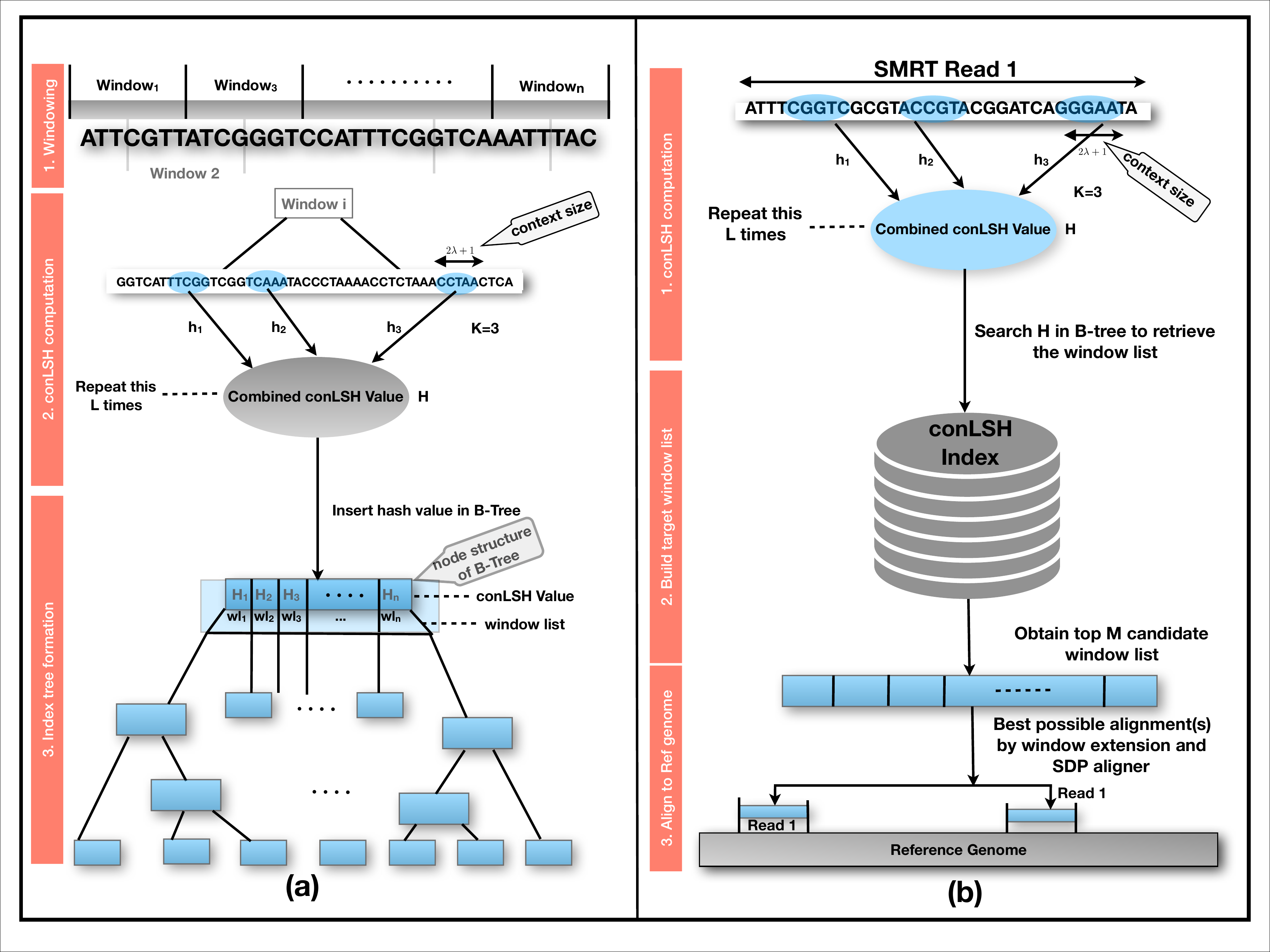}
\caption{(a) describes the three different stages: Windowing, conLSH computation and Index tree formation of conLSH-indexer. conLSH-aligner computes the hash value for the reads and uses the index tree build by conLSH-indexer to map reads back to the reference genome. The worflow of conLSH-aligner is depicted in Fig.(b), which goes through the phases of conLSH computation, Build target window list, and Alignment to reference genome.}
\label{Workflow of conLSH Indexer and Aligner}
\end{figure}

 Hashing is a popular approach of indexing objects for fast retrieval. In conLSH, sequences are hashed together if they share similar contexts. The probability of collision of two sequences is proportional to the number of common contexts between them. The \emph{context}s play an important role to group sequences having noisy bases like PacBio data.  The workflow of indexing reference genome and alignment of SMRT reads using conLSH, is portrayed in the Figure \ref{Workflow of conLSH Indexer and Aligner}. At the indexing phase, the reference genome is virtually split into several overlapping windows. For each window, a set of conLSH values has been computed and stored as a B-Tree index (for details refer Supplementary Note 1) to facilitate logarithmic index search. The aligner hashes the SMRT reads using the same conLSH functions and retrieves the window-list from B-Tree index that resulted into the same hash values as the read. This forms the list of candidate sites for possible alignment after extension. Finally, for each read, the best possible alignment(s) are obtained by Sparse Dynamic Programming (SDP). \\
 The combined hashed value, as shown in Fig. \ref{Workflow of conLSH Indexer and Aligner}, captures contexts from $K$ (\emph{concatenation factor}) different locations, where each context is of size $2\times \lambda+1$, ($\lambda$ is the \emph{context factor}, Refer Methods Section). This increases the seed length to resolve repeats and thereby prevents false positive targets. Therefore, even if a kmer is repetitive, conLSH aligner can map it properly with the help of its contexts. The problem occurs when the repeat regions are longer than the reads. However, this is very unlikely due to the increased read length of SMRT data. The number of hash tables ($L$), context factor ($\lambda$) and concatenation factor ($K$) play an important role to make the aligner work at its best. An increase in $\lambda$ and $K$, will make the aligner more selective. Then, $L$ should be sufficiently large to ensure that the similar sequences are hashed together in at least one of the $L$ hash tables \cite{LSH_new}. It may happen that a longer read could not be mapped to the reference genome by an end to end alignment. In that case, conLSH has a provision of chimeric read alignment where a read is split and rehashed to find the candidate sites for each split, and aligned accordingly.

%The reference genome and SMRT reads are hashed using the same set of suitably designed conLSH functions. The reads hashed together with reference genome in the conLSH table, give an insight of the target regions for possible extension and alignment of SMRT reads. A Sparse Dynamic Programming based approach is used to align the reads to the targeted sites.

\section*{Results}

We have conducted exhaustive experiments on both simulated and real PacBio datasets to assess the performance of our proposed conLSH based aligner. The result is demonstrated in three categories: i)Performance on Reference Genome Indexing, ii)Results of SMRT read Alignments and iii)Robustness of conLSH aligner to different parameter settings. Five real (\emph{E.coli, A.thaliana, O.sativa, S.cerevisiae, H.sapiens}) and  four simulated  SMRT datasets were used to benchmark the performance of conLSH and four other state-of-the-art aligners, rHAT\cite{liu2015rhat}, BLASR\cite{chaisson2012mapping}, BWA-MEM\cite{li2013aligning} and BWA-SW\cite{li2010fast}. A detailed description of the datasets along with their accession link is included in Supplementary Table 1. A brief manual for different working parameters and a \emph{Quick Start} guide can be found at Supplementary Note 3. The experiment has been conducted on an Intel® Core™ i7-6200U CPU @ 2.30GHz $\times$ 8(cores), 64-bit machine with 16GB RAM.

\subsection*{Performance on Reference Genome Indexing}

Indexing of reference genome is the first step of the pipeline. The index file generated in this phase tries to encompass the pattern lies in the entire genome, which is later used by the aligner to map the reads back to the reference. Table \ref{Table1} portrays the performance of conLSH-indexer in comparison with two other state-of-the-art methods, rHAT and BWA. BLASR is not listed here, because it works better with BWT-index\cite{chaisson2012mapping}. We executed rHAT for two different parameter settings, rHAT1 with -l 11, -w 1000, -m 5, -k 13 (default settings) and rHAT2 with -l 8, -w 1000, -m 5, -k 15 (settings similar to conLSH).   
\begin{table*}[h]
\centering{}
\footnotesize
\begin{tabular}{c|cc|cc|cc|cc|cc}
\hline
\\
\textbf{Dataset}  &\multicolumn{2}{c}{\emph{E.coli}}&\multicolumn{2}{c}{\emph{H.sapiens}}&\multicolumn{2}{c}{\emph{A.thaliana}}&\multicolumn{2}{c}{\emph{S.cerevisiae}}&\multicolumn{2}{c}{\emph{O.sativa}}\\
\cline{2-11}
&Size(MB)&Time(S)&Size(MB)&Time(S)&Size(MB)&Time(S)&Size(MB)&Time(S)&Size(MB)&Time(S)\\
\hline
conLSH	&	\textbf{1.38}	&	\textbf{0.01}	&	\textbf{724.32}	&	\textbf{49}	&	\textbf{29.87}	&	\textbf{1}	&	\textbf{3.03}	&	\textbf{0.01}	&	\textbf{93.18}	&	\textbf{7}\\
rHAT1	&	290	&	4	&	11755.34	&	785	&	742.84	&	35	&	316.91	&	3	&	1744.08	&	122\\
rHAT2	&	4317	&	49	&	15806.77	&	1166	&	4770.16	&	115	&	4343.47	&	50	&	5804	&	253\\
BWA	&	9.7	&	8.7	&	4523	&	3904	&	209	&	114.7	&	21.3	&	10.16	&	652.4	&	543.3\\
\hline
\hline
\end{tabular}
\par
\caption{The time required to produce the reference index and the index file size by three different methods for \emph{E.coli, H.sapiens, A.thaliana, S.cerevisiae,} and \emph{O.sativa} genomes} \label{Table1}
\end{table*}

It can be observed from Table \ref{Table1} that conLSH is the fastest one to produce the reference index,  and that is too of the least size. conLSH yields a reduction of $94\%$ in index size, while saving $93.7\%$ of the time for \emph{H.sapiens} genome, in comparison to rHAT (default). The gain in time and index size attained by conLSH increases for rHAT2. BWA, on the other hand, produces indexes of size smaller than that of rHAT, but at the cost of time. 

\subsection*{Results of SMRT Read Alignments}

We have used the PacBio P5/C3 release datasets respectively from \emph{H.sapiens, E.coli, S.cerevisiae, A.thaliana} and \emph{O.sativa} (Please refer Supplementary Table 1 for details) genomes to extensively study the performance of proposed conLSH-aligner for noisy and long SMRT reads. Table \ref{Table2} summarizes the results for \emph{H.sapiens} dataset by conLSH, rHAT, BWA-MEM, BWA-SW and BLASR, in terms of time taken, percentage of the reads aligned and the peak memory footprint. The working parameters of BWA-MEM, BWA-SW and BLASR, as found to be appropriate\cite{liu2015rhat} for  PacBio data, is mentioned in the table. conLSH operating at a default settings of -l 8, -w 1000, -m 5, - $\lambda$  2, -K 2, -L 1, generates patterns of length 16 (context size=$2\lambda+1$=5 and concatenation factor (K)=2) assuming a gap of 3 bases between the contexts. To make rHAT comparable with conLSH, a settings (rHAT2) with longer kmer (-k 15) has been used rather than the default kmer size of 13 (rHAT1).

\begin{table*}[h]
\centering{}
\footnotesize
\begin{tabular}{c|c|c|c|c}
\hline
\\
\textbf{Methods} & \textbf{Working Parameters} & \textbf{Time Taken (Sec)}& \textbf{\% of Reads Aligned} & \textbf{Peak Memory Footprint (MB)}\\

\hline
\hline
rHAT1	&	-l 11, -w 1000, -m 5, -k 13	&	210839	&	99.8	&	14562\\
rHAT2	&	-l 8, -w 1000, -m 5, -k 15	&	667280	&	\textbf{100}	&	15746\\
conLSH	&	-l 8, -w 1000, -m 5, - $\lambda$  2, -K 2, -L 1	&	\textbf{159888}	&	99.8	&	\textbf{4327}\\
BLASR	&	BWT	&	724304	&	99.7	&	8100\\
BLASR	&	SA	&	848602	&	99.8	&	14700\\
BWA-MEM	&	-x PACBIO	&	372782	&	99.7	&	5200\\
BWA-SW	&	-b 5, -q 2, -r 1, -z 20	&	1145150	&	99.76	&	7100\\

\hline
\hline
\end{tabular}
\par
\caption{Performance of rHAT, conLSH, BLASR, BWA-MEM and BWA-SW for real PacBio H.sapiens P5/C3 release } \label{Table2}
\end{table*}

\begin{figure*}[ht]
        \centering
        \begin{subfigure}[t]{0.3\textwidth}
            \centering
            \includegraphics[scale=0.4]{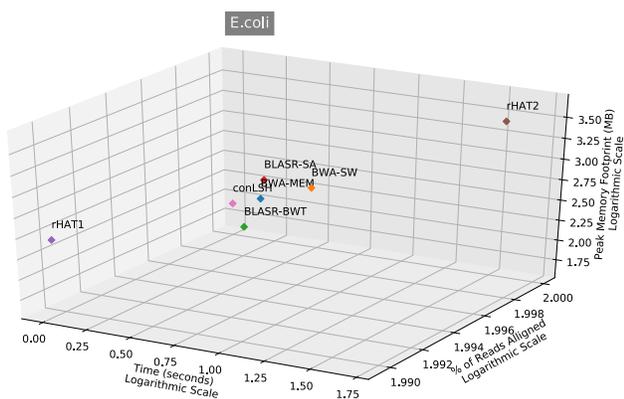}
            \caption[]%
            {{\small E.coli PacBio Dataset}}    
            \label{fig:ecoli}
        \end{subfigure}
        \hfill
        \begin{subfigure}[t]{0.5\textwidth}  
            \centering 
            \includegraphics[scale=0.4]{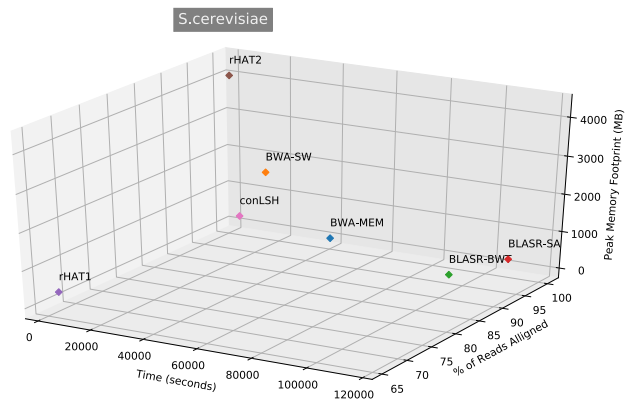}
            \caption[]%
            {{\small S.cerevisiae PacBio Dataset}}    
            \label{fig:cere}
        \end{subfigure}
        \\
        %\vskip\baselineskip
        \hfill
        \begin{subfigure}[t]{0.45\textwidth}   
            \centering 
            \includegraphics[scale=0.4]{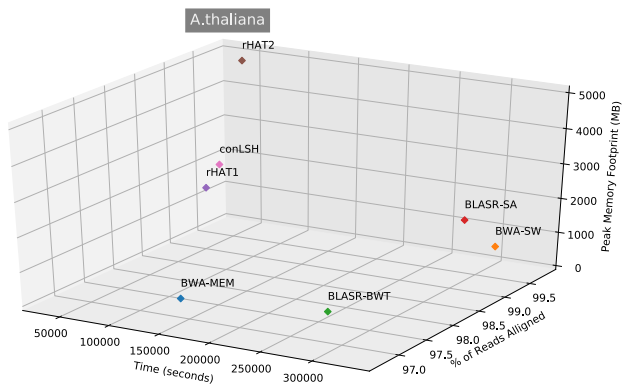}
            \caption[]%
            {{\small A.thaliana PacBio Dataset}}    
            \label{fig:thaliana}
        \end{subfigure}
        \hfill
        %\vskip\baselineskip
        \begin{subfigure}[t]{0.50\textwidth}   
            \centering 
            \includegraphics[scale=0.4]{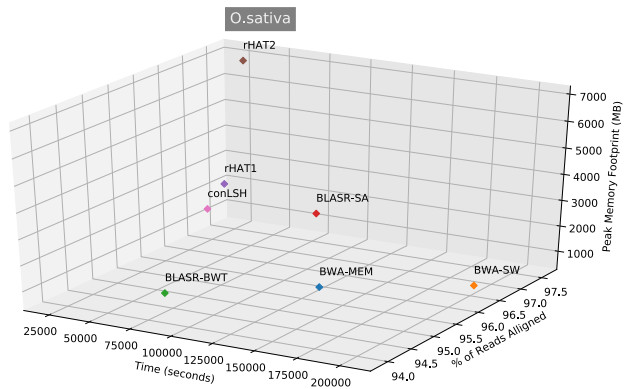}
            \caption[]%
            {{\small O.sativa PacBio Dataset}}    
            \label{fig:sativa}
        \end{subfigure}
       
        \caption[  ]
        {\small Peak Memory footprint, \% of Reads Aligned and Time Taken by rHAT1, rHAT2, conLSH, BLASR-BWT, BLASR-SA, BWA-MEM, BWA-SW for four real PacBio datasets of \emph{E.coli, S,cerevisiae, A.thaliana, O.sativa} genomes} 
        \label{distribution}
    \end{figure*}

Table \ref{Table2} shows that conLSH aligns $99.8\%$  reads of the real \emph{H.sapiens} dataset in least time, producing a time gain of $24.2\%$ and $76\%$ over rHAT1 and rHAT2 respectively. The smallest run-time memory requirement of 4.3GB, which is $70.3\%$ less than that of rHAT1, makes conLSH an attractive choice as SMRT aligner. The longer PacBio reads provide better repeat resolution by facilitating larger kmer match. However, the huge memory requirement of 15.7GB, even at a kmer size of 15, makes rHAT infeasible to work with longer stretches. Though, BWA-MEM , BWA-SW and BLASR yields lesser memory footprint in comparison to rHAT, the performance is limited by the alignment speed. A similar scenario is reflected in Figure \ref{distribution}, when studied with other PacBio datasets of \emph{E.coli, S.cerevisiae, A.thaliana} and \emph{O.sativa}. conLSH has been found to consistently outperform the state-of-the-art methods in terms of memory consumption, while maintains a high throughput at the same time. 
 
\newpage
\begin{figure}[H]
\centering
\includegraphics[scale=.5,trim=4 4 4 4,clip]{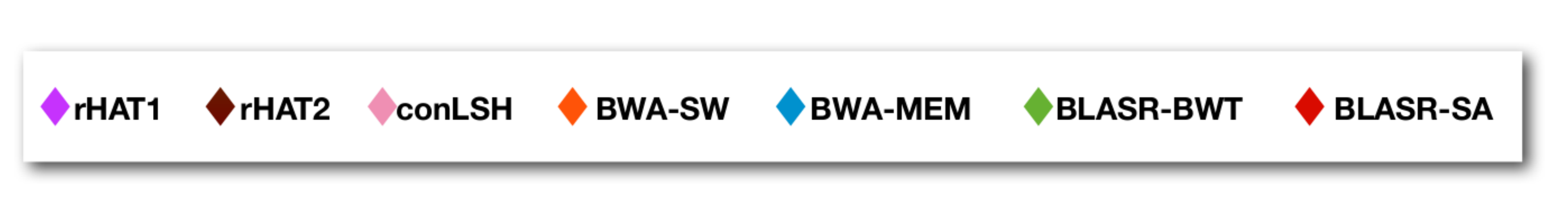}
\end{figure}

  \subsubsection*{Alignment on Simulated Datasets}
In this section for simulated datasets, we have restricted our study to rHAT and conLSH only, as rHAT has been found to work better\cite{liu2015rhat} than BWA and BLASR in terms of alignment speed and quality for SMRT reads. We used four different simulated datasets, \emph{E.coli, S,cerevisiae, A.thaliana} and \emph{D.melanogaster}, originally generated to test the performance of rHAT\cite{liu2015rhat} using PBSim\cite{pbsim} with 1x coverage having average read length of 8000bp and 15\% error probability (12\%  insertions, 2\% deletions and 1\% substitutions) resembling the PacBio P5/C3 dataset (Please refer Supplementary Table 1 for details).  
  \begin{table*}[h]
\centering{}
\footnotesize
\begin{tabular}{c|c|c|c|c|c}
\hline
\\
\textbf{Dataset}&\textbf{Methods} & \textbf{Working Parameters} & \textbf{Time Taken (Sec)}& \textbf{\% of Reads Aligned} & \textbf{Peak Memory Footprint (MB)}\\

\hline
\hline
\multirow{ 3}{*}{\emph{E.coli\_sim}}
& rHAT1	&	-l 11, -w 1000, -m 5, -k 13	&	6	&	94.8	&	294\\
& rHAT2	&	-l 8, -w 1000, -m 5, -k 15	&	59	&	100	&	4225\\
& conLSH	&	-l 8, -w 1000, -m 5, - $\lambda$  2, -K 2, -L 1	&	15	&	100	&	\textbf{85}\\
\hline
\multirow{ 3}{*}{\emph{S.cerevisiae\_sim}}
& rHAT1	&	-l 11, -w 1000, -m 5, -k 13	&	9	&	100	&	326\\
& rHAT2	&	-l 8, -w 1000, -m 5, -k 15	&	85	&	100	&	4259\\
& conLSH	&	-l 8, -w 1000, -m 5, - $\lambda$  2, -K 2, -L 1	&	45	&	100	&	\textbf{156}\\
\hline
\multirow{ 3}{*}{\emph{A.thaliana\_sim}}
& rHAT1	&	-l 11, -w 1000, -m 5, -k 13	&	89	&	100	&	849\\
& rHAT2	&	-l 8, -w 1000, -m 5, -k 15	&	372	&	100	&	4781\\
& conLSH	&	-l 8, -w 1000, -m 5, - $\lambda$  2, -K 2, -L 1	&	337	&	99.9	&	\textbf{574}\\
\hline
\multirow{ 3}{*}{\emph{D.melanogaster\_sim}}
& rHAT1	&	-l 11, -w 1000, -m 5, -k 13	&	139	&	100	&	1054\\
& rHAT2	&	-l 8, -w 1000, -m 5, -k 15	&	223	&	98	&	4987\\
& conLSH	&	-l 8, -w 1000, -m 5, - $\lambda$  2, -K 2, -L 1	&	395	&	99	&	\textbf{705}\\

\hline
\hline
\end{tabular}
\par
\caption{Comparative study of Memory Requirement, \% of Reads Aligned and Time Taken by rHAT and conLSH for four different simulated datasets of \emph{E.coli, S,cerevisiae, A.thaliana} and \emph{D.melanogaster} genomes respectively }
 \label{Table3}
\end{table*}

The Table \ref{Table3} describes the performance of rHAT1 (default settings), rHAT2 (settings similar to conLSH) in comparison with the proposed conLSH-aligner for simulated datasets. It is evident that conLSH has the lowest memory footprint for all the datasets. With similar kmer size, rHAT2 consumes $92\%$ more space, on average, than conLSH. The speed of rHAT aligner is achieved at the cost of huge memory requirement. The exhaustive hashing makes the search faster, but increases the index size. In spite of that, conLSH works faster than rHAT2 for most of the datasets, as can be seen from Table \ref{Table3}, except \emph{D.melanogaster}. Though, rHAT1 works reasonably good in terms of memory footprint and alignment speed for these simulated datasets, it fails to take the advantage of longer SMRT reads due to the small size of kmers. An exhaustive study on simulated datasets for different working parameters of conLSH can be found at Supplementary Table 2.  
\subsubsection{Quality of Alignment}

\begin{figure*}[h]
        \centering
        \begin{subfigure}[t]{0.4\textwidth}
            \centering
            \includegraphics[scale=0.38]{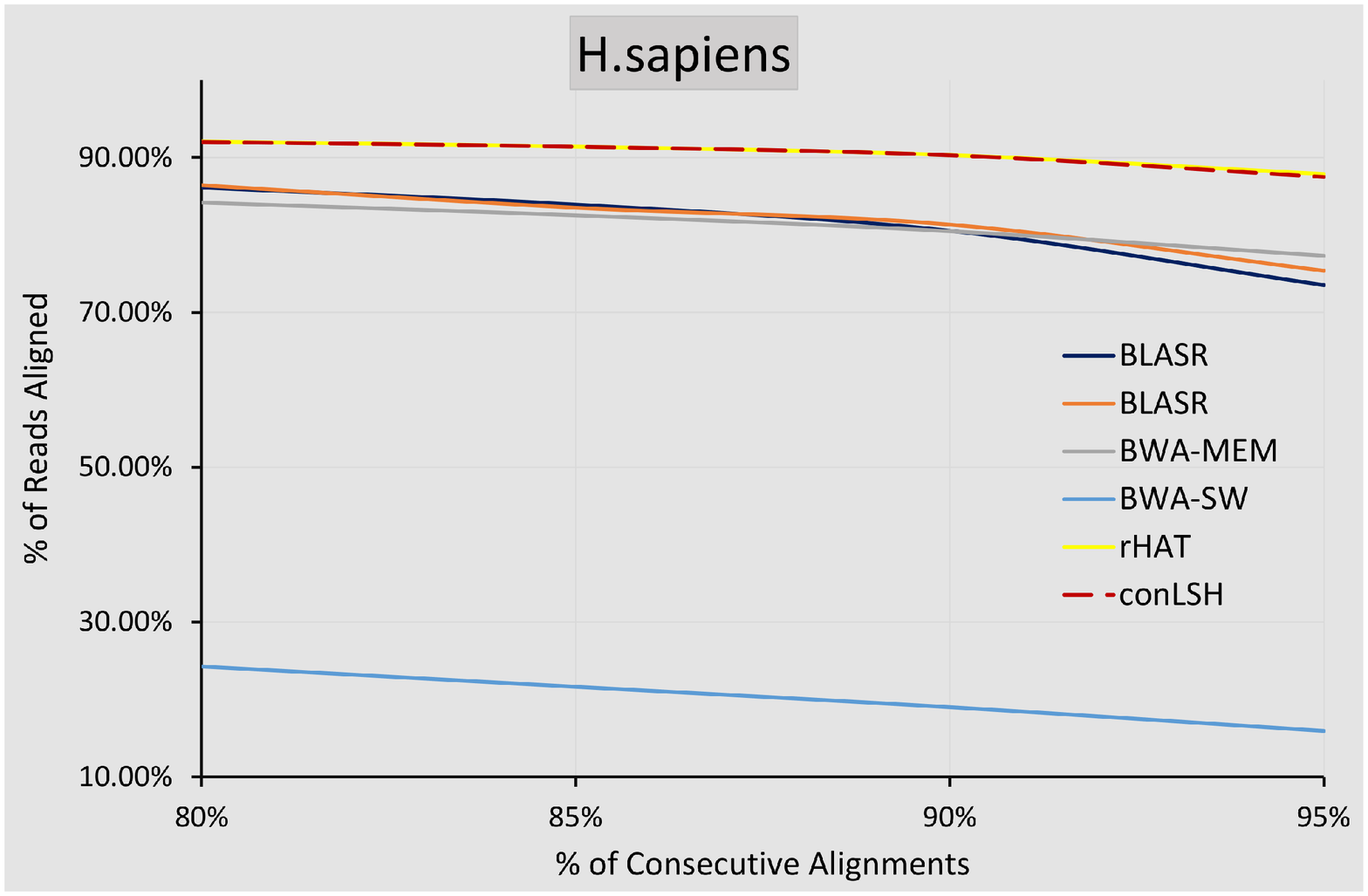}
            \caption[]%
            {{\small H.sapiens PacBio Dataset}}    
            \label{fig:ecoli}
        \end{subfigure}
        \hfill
        \begin{subfigure}[t]{0.47\textwidth}  
            \centering 
            \includegraphics[scale=0.38]{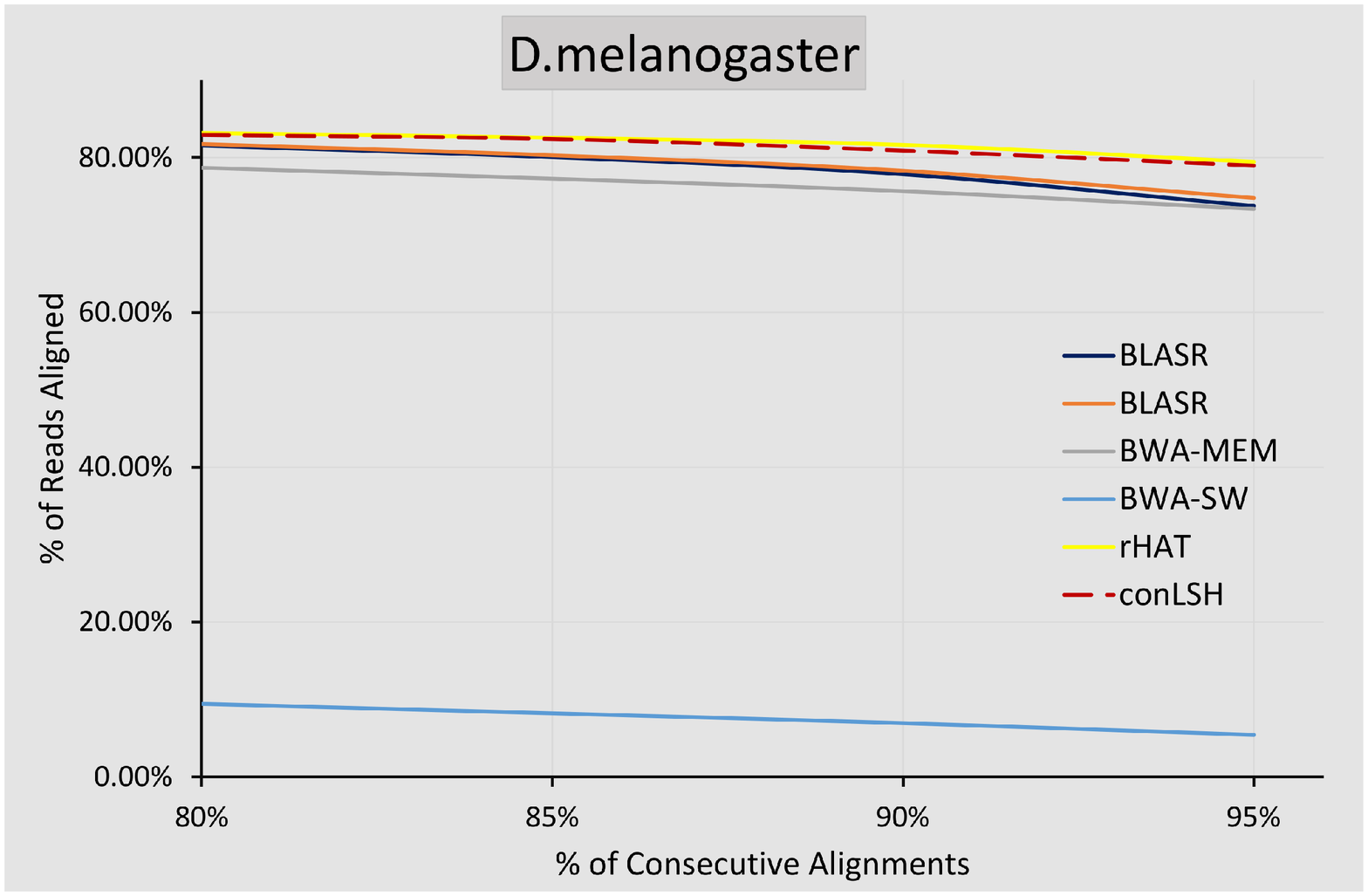}
            \caption[]%
            {{\small D.Melanogaster PacBio Dataset}}    
            \label{fig:cere}
        \end{subfigure}

        \caption[  ]
        {\small Reads aligned consecutively by different methods for the datasets, \emph{H.sapiens} and \emph{D.melanogaster}} 
        \label{consecutiveness}
    \end{figure*}

To assess the quality of the alignments, we have conducted an experiment to measure the number of reads aligned consecutively by the respective aligners. Figure \ref{consecutiveness} plots the \% reads aligned with a specific amount of continuity for real \emph{H.sapiens} and {D.melanogaster} datasets. It can be observed that conLSH has $92.02\%$ of the aligned reads with $80\%$-consecutiveness. The amount is almost similar for rHAT and gradually decreases for BWA-MEM, BLASR and BWA-SW. This this due to the fact that after shortlisting the target sites for read mapping, conLSH adopts the same procedure of window expansion and SDP based alignment as done by rHAT. This makes the alignment quality of rHAT and conLSH quite similar, which is reflected in Figure \ref{consecutiveness}. 

%\begin{figure}[ht]
%\centering
%\includegraphics[scale=0.3]{plot.eps}
%\caption{Consecutively of the Aligned Reads}
%\label{Workflow of conLSH Indexer and Aligner}
%\end{figure}

    \subsection*{Study of Robustness for different parameter values}
    To assess robustness of the proposed method, we have studied the performance of conLSH for different working parameters. The result of conLSH indexer and aligner on subread\_m130928\_232712\_42213 of Human genome for different values of concatenation factor ($K$), context size ($2\lambda+1$) and number of hash tables ($L$), are comprehensively reported in Table \ref{Table4}. It is clear that the size of the reference index is directly proportional with the number of hash tables ($L$). The parameters $K$ and $\lambda$, on the other hand, decide the kmer size for pattern matching and therefore, have a strong correlation with the memory footprint of the aligner. The parameters, $K$ and $\lambda$ together control the size of the BTree index in memory for storing the hash values and the corresponding window lists. An increase in $K$ and $\lambda$ makes the aligner more selective in finding the target sites and consequently, increases the sensitivity. To ensure a desired level of throughput, $L$ should be sufficiently large so that the similar sequences are hashed together into at least one of the $L$ hash tables. Maintaining a trade-off between the memory footprint and throughput, a settings of $K=2, \lambda=2, L=1$ has been chosen as the default (marked as bold in Table \ref{Table4}). However, it is needless to mention that the performance of conLSH does not change severely with the change of parameters. Please note that, with $\lambda=0$, the context size becomes 1 and the algorithm works as normal LSH based method without looking for any contexts to match. It can be concluded from Table \ref{Table4} that for a particular value of $K$ and $L$, an increase in the context size enhances the sensitivity of the aligned reads as well as the throughput of the aligner. A similar study with different working parameters for four other real datasets has been included in Supplementary Table 3.

  \begin{table*}[h]
\centering{}
\footnotesize
\begin{tabular}{c|c|c|cc|ccc}
\hline
\textbf{Concatenation}&\textbf{Context Size} & \textbf{L}  &\multicolumn{2}{c}{\textbf{conLSH\_indexer}}& \multicolumn{3}{c}{\textbf{conLSH\_aligner}}\\
\textbf{Factor (K)} & \textbf{$2\times \lambda +1$}& &\textbf{Time(s)}&\textbf{Hash\_size(MB)}&\textbf{Time(s)}&\textbf{\%of Reads Aligned}& \textbf{Peak Memory (MB)}\\
\hline
\hline

2	&	1	&	1	&	46	&	724.32	&	341	&	97.9	&	3811	\\
2	&	1	&	2	&	57	&	1448.65	&	424	&	95.7	&	4542	\\
\hline
2	&	3	&	1	&	49	&	724.32	&	391	&	98.3	&	3813	\\
2	&	3	&	2	&	56	&	1448.65	&	390	&	98.4	&	4521	\\
\hline
\textbf{2}	&	\textbf{5}	&	\textbf{1}	&	\textbf{49}	&	\textbf{724.32}	&	\textbf{444}	&	\textbf{99}	&	\textbf{4327}	\\
2	&	5	&	2	&	55	&	1448.65	&	505	&	99.1	&	5050	\\
\hline
3	&	1	&	1	&	79	&	724.32	&	439	&	99	&	3799	\\
3	&	1	&	2	&	62	&	1448.65	&	441	&	99.1	&	4518	\\
\hline
3	&	3	&	1	&	49	&	724.32	&	452	&	99.1	&	3932	\\
3	&	3	&	2	&	58	&	1448.65	&	468	&	99.2	&	4651	\\

\hline
\hline
\end{tabular}
\par
\caption{Performance of conLSH with the change of K, L and $\lambda$ for subread\_m130928\_232712\_42213 of \emph{H.sapiens} dataset} \label{Table4}
\end{table*}   
\section*{Discussion}

In this article, we have conceived the idea of \emph{context based} locality sensitive hashing and deployed it to align noisy and long SMRT reads. The strategy of grouping sequences based on their contextual similarity not only makes the aligner work faster, but also enhances the sensitivity of the alignment. The result is evident from the experimental study on different real and simulated datasets.  It has been observed that larger seeds are helpful for aligning noisy and repetitive sequences \cite{chaisson2012mapping}. The combined hash value, as obtained from the contexts of different locations of the sequence stream, makes conLSH work more accurately in comparison to rHAT, a recently developed state-of-the-art method. Having a 32-bit hash value, the maximum possible kmer size of rHAT is 16 (as, $2^{32}=4^{16}$). The \emph{PointerList} of rHAT stores links to all possible $4^k$ hash values incurring a huge memory cost. As a result, rHAT has a memory requirement of 13.7GB and 17.48GB with kmer size of 13 and 15 respectively for \emph{H.sapiens} dataset. The exponential growth of RAM space with the increase of kmer size prohibits rHAT from using larger seeds. Moreover, rHAT generates 256 auxiliary files during indexing of reference genomes. This huge space constraint restricted the utility of rHAT and demanded for a fast as well as memory efficient solution. conLSH is an effort in this direction. It efficiently reduces the search space by storing the hash values and the corresponding window lists dynamically on demand.   B-tree index of the proposed aligner also facilitates efficient disk retrieval for huge genome, if could not be accommodated in memory. The selection of context size ($\lambda$) adds extra flexibility to the aligner. An increase in $\lambda$ and $K$ increases the seed size and makes the aligner more sensitive. This mode of operation is specially suitable for aligning repetitive stretches. Therefore, conLSH can serve as a drop-in replacement for current SMRT aligners.\\

conLSH also has some limitations to mention. The Sparse Dynamic Programming (SDP) based heuristic alignment is time consuming and may lead to false positive alignments \cite{liu2015rhat}. As we have inherited the SDP based target site alignment from rHAT, the shortcomings of it are also there in conLSH. Though, the default parameter settings work reasonably good for most of the datasets studied, conLSH requires a understanding of context factor and LSH parameters in order to be able to use it. We have future plan to automate the parameter selection phase of the algorithm by studying the sequence distribution of the dataset. conLSH can be further enhanced by incorporating parallelization taking into account of the quality scores of the reads.

\section*{Methods}
\label{method}
Nearest neighbor search is a classical problem of pattern analysis, where the task is to find the point nearest to a query from a collection of $d$-dimensional data points. A less stringent version of exact nearest neighbor search is to find the $R$-nearest neighbors, in which the algorithm  finds all the neighbors within a certain distance $R$ of the query. It can be formally defined as follows:\\
\theoremstyle{definition}
\newtheorem{definition}{Definition}
\emph{
\begin{definition}{$R$-Nearest Neighbor Search: \cite{LSH_new}} \\
Given a set of $n$ points $X$= $\{x_1, x_2,\dots, x_n\}$ in $d$-dimensional space $\mathbb{R}^d$   and a query point $y$, it reports all the points from the set $X$ which are at most $R$ distance away from $y$, where $R>0$.
\end{definition}}

Due to the ``Curse of Dimensionality" the performance of most of the partitioning based algorithms, like kd-trees, degrades for higher dimensional data. However, there are few exceptions \cite{muja2014scalable} where the authors have shown good results with priority search k-means tree based algorithm. A solution to this problem is approximate nearest neighbor search. These approximate algorithms return, with some probability, all points from the dataset which are at most $c$ times $R$ distance away from the query object. The standard definition of $c$-approximate nearest neighbor search is:\\
\emph{
\begin{definition}{$c$-approximate $R$-near Neighbor Search: \cite{LSH_new}} \\
Given a set of $n$ points $X$= $\{x_1, x_2,\dots, x_n\}$ in $d$-dimensional space $\mathbb{R}^d$  and a query point $y$, it reports all the points $x$ from the set $X$ where $d(y,x)\le cR$ with probability $(1-\delta)$, $\delta>0$, provided that there exists an $R$-near neighbor of $y$ in $X$, where $R>0$ and $c>1$.
\end{definition}}

The advantage of using approximate algorithms is that they work very fast and the reported approximate neighbors are almost as good as the exact ones \cite{LSH_new}. Locality Sensitive Hashing based approximate search algorithms are very popular for high dimensional data. LSH works on the principle that the probability of collision of two objects is proportional to their similarity. Therefore, after hashing all the points in the dataset, similar points will be clustered in the same or localized slots of the hash tables. LSH is defined as: \cite{LSH_old98}\\

\begin{definition}{Locality Sensitive Hashing:} \\
A family of hash functions $\mathcal{H}:\mathbb{R}^d\to U$ is called $(R,cR,P_1,P_2)$-sensitive if $\forall x,y\in\mathbb{R}^d$
\begin{itemize}
 \item
  if $\parallel x-y\parallel\leq R$, then $Pr_{\mathcal{H}}[h(x)=h(y)]\geq P_1$\\
  \item
  if $\parallel x-y\parallel\geq cR$, then $Pr_{\mathcal{H}}[h(x)=h(y)]\leq P_2$, where, $P_1>P_2$.
  \end{itemize}
\end{definition}

We have proposed the following definitions and theories to develop the framework of Context based Locality Sensitive Hashing(conLSH).\\

 \begin{definition}{Context based Locality Sensitive Hashing (conLSH):} \\
A family of hash functions $\mathcal{H^*}$ is defined as $\mathcal{H^*}:\mathbb{R}^{d+1}\to U$ where the additional dimension denotes the context factor. $\mathcal{H^*}$ is called $(R,cR,\lambda,P^*_1,P^*_2)$-sensitive if $\forall x,y\in\mathbb{R}^{d}$
\begin{itemize}
 \item
  if $\parallel x^\lambda-y^\lambda\parallel\leq R$, then $Pr_{\mathcal{H^*}}[h^*(x,\lambda)=h^*(y,\lambda)]\geq P^*_1$\\
  \item
  if $\parallel x^\lambda-y^\lambda\parallel\geq cR$, then $Pr_{\mathcal{H^*}}[h^*(x,\lambda)=h^*(y,\lambda)]\leq P^*_2$, where, $P^*_1>P^*_2$.
  \end{itemize}
\end{definition}

Here $h^*(x,\lambda)$ is the context based locality sensitive hash function and $x^\lambda$ is the point $x$ associated with a context of size $(2\lambda+1)$, where $\lambda$ is the \emph{context factor}. Let $x_1, x_2, \dots, x_n$ and $y_1, y_2, \dots, y_m$ are two sequences on which conLSH is being applied.  Then, $h^*(x_i,\lambda)$ is actually a composition of $(2\lambda+1)$ standard LSH functions as defined below:
 \begin{definition}{$h^*(x_i,\lambda)$:} \\
$h^*(x_i,\lambda)=h(x_{(i-\lambda)}) \circ h(x_{(i-(\lambda-1)}) \circ \dots \circ h(x_i) \circ h(x_{(i+1)}) \circ \dots \circ h(x_{(i+\lambda)})$, where $1 \le i \le n $.
\end{definition}

From now onwards, the context based LSH function $h^*(x_i,\lambda)$ is shortened as $h_i^\lambda$. In the next subsection, we introduce the $K$ and $L$ parameter values in connection with the concatenation of hash functions to amplify the gap between $P_1$ and $P_2$.

\subsection{Gap Amplification} \label{gap_sec}
As the gap between the probability values $P_1$ and $P_2$ is very small, several locality sensitive hash functions are usually  concatenated to obtain the desired probability of collision. This process of concatenation of hash functions to increase the gap is termed as ``gap amplification". Let there be $L$ hash functions, ${g_1,g_2,\dots, g_L}$, such that  $g_j$ is the concatenation of $K$ randomly chosen hash functions like, $g_j=(h_{1,j}^\lambda,h_{2,j}^\lambda, \dots, h_{K,j}^\lambda)$ for $1 \leq j \leq L$. Therefore, the probability that $g_j(x)=g_j(y)$ is at least $P_1^{(2\lambda+1)K}$. Each point, $x$, from the database is placed into the proper slots of the $L$ hash tables using the values $g_1(x), g_2(x), \dots, g_L(x)$ respectively. Later when the query, $y$ comes, we search though the buckets $g_1(y), g_2(y), \dots, g_L(y)$.

Note that larger values of $K$ lead to larger gap between the LSH probabilities. As a consequence, the hash functions become more conservative in estimating the similarity between points. In that case, $L$ should be large enough to ensure that  similar points collide with the query at least once. The context factor $\lambda$ should also be chosen carefully because if it is too large it will overestimate the context, thereby making hash functions more stringent. A detailed theoretical discussion on the choice of the parameters $K, L,$ and $\lambda$ is included in Supplementary Note 2.

\subsection{Algorithm and Analysis}

\renewcommand{\algorithmicrequire}{\textbf{Input:}}
\renewcommand{\algorithmicensure}{\textbf{Output:}}

\begin{algorithm}
\caption{Indexing of Reference Genome using conLSH}
\label{indexalgo}
\begin{algorithmic}[1]
\begin{multicols}{2}
\REQUIRE Let $R$ be the Reference Genome\\
		$K$,$L$,$\lambda$ are the conLSH parameters
		 \\context size=$2\lambda+1$\\
		$w$=window size (default=2000bp)
		
\ENSURE  conLSH index of $R$ is stored in B-Tree

\STATE $R$ is split into $n$ windows, $(W_1,W_2,\dots, W_n)$, each of size $w$, having overlap of $w/2$ bp.\\
      $|W_i|=w$, for $1\le i\le n$
\STATE $i\gets 1$
  \WHILE{$i \leq n$}
   \STATE conLSH value will be computed for window $W_i$
   \STATE $j\gets 1$
    \WHILE{$j \leq L$}
    \STATE  Place $W_i$ in the $g_j(W_i)$ th slot of the $j$th hash table, where $g_j(W_i)=(h_{1,j}^{\lambda} \circ h_{2,j}^{\lambda} \circ \dots \circ h_{K,j}^{\lambda})$ \\
    $h_{k,j}^{\lambda}$ is randomly chosen from $\mathcal{H^*}$ along with a context of size $(2\lambda+1)$, $1\le k \le K$.
    \STATE Insert the window position and the corresponding conSH value in B-Tree Index.
    \STATE $j \gets j+1$
    \ENDWHILE 
    \STATE $i \gets i+1$
  \ENDWHILE
\STATE The B-Tree with all window positions along with their conLSH values forms the reference index.
\end{multicols}
\end{algorithmic}

\end{algorithm}

\begin{algorithm}
\caption{Alignment of PacBio Reads to the Reference Genome}
\label{alignalgo}
\begin{algorithmic}[1]
\begin{multicols}{2}
\REQUIRE Long, noisy SMRT reads and the alignment scores (match, mismatch and gap penalties)
\ENSURE  Best possible alignment(s) for each read in the reference genome 

	\FOR {each PacBio read $r$}
   \STATE Compute the conLSH values for the $L$ hash tables
   \STATE $j\gets 1$
   \WHILE{$j\leq L$}
   \STATE Place r in $g_j(r)$ th slot in the $j$th hash table, where $g_j$ is the same conLSH function defined in Algorithm \ref{indexalgo}.
   \STATE Search B-Tree index for $g_j(r)$ to retrieve the window list that are mapped with the same hash value.  
   \ENDWHILE

   \STATE Obtain $M$ most frequent candidate windows from the window list.
   \FOR{each candidate window $C_i$, $1\le i \le M$}
   \STATE $C_i$ window is further extended to obtain the probable alignment site as done in rHAT\cite{liu2015rhat}.
   \STATE Finally, the SDP-based heuristic alignment approach has been inherited from rHAT, where a directed acyclic graph (DAG) is build to find the mapping.  
   \ENDFOR
   \STATE If the read $r$ could not be successfully aligned to any of the probable candidate sites, the algorithm tries to find chimeric read alignment. It splits a read into several fixed sized segments and the same procedure of conLSH computation and target location search is applied for each of the segments.
    \ENDFOR
\end{multicols}
\end{algorithmic}

\end{algorithm}

\textbf{Space Complexity}

The space consumed by the algorithm is dominated by the hash tables and the data stored in it. Let there be a total of $n$ reads to be placed into $L$ different hash tables.
Therefore, the space requirement is
=$\mathcal{O}(L \times n)$\\
=$\mathcal{O}(n^\rho \cdot \ln\frac{1}{\delta} \cdot n)$, See Supplementary Note 2.\\
=$\mathcal{O}(\ln\frac{1}{\delta} \cdot n^{\rho+1})$=$\mathcal{O}(n^{\rho+1})$\\

\textbf{Indexing Time:}\\
For each sequence, $L$ different hash values are computed for $L$ hash tables. Each hash function is composed of $(2\lambda+1)\times K$ unit hash functions. Therefore, the time required to map a single read to its suitable positions in $L$ different hash tables is $\mathcal{O}((2\lambda+1)\cdot K \cdot L)$\\
Therefore, the total time requirement for $n$ sequences is $\mathcal{O}((2\lambda+1)\cdot K \cdot L\cdot n)$\\
=$\mathcal{O}((2\lambda+1) \cdot (\frac{1}{2\lambda+1})\frac{\ln n}{\ln \frac{1}{P_2}} \cdot L \cdot n)$, See  Supplementary Note 2\\
=$\mathcal{O}(\frac{\ln n}{\ln \frac{1}{P_2}} \cdot n^\rho \cdot \ln\frac{1}{\delta} \cdot n)$=$\mathcal{O}( \frac{n^{\rho+1} \cdot \ln n}{\ln \frac{1}{P_2}})$\\

\vspace{30pt}

\textbf{Query Time:}\\
The search for target windows from the sequences that are hashed together with the query, is usually stopped after reporting $L^\prime=3L$ candidates. This ensures a reasonably good result with constant error probability\cite{LSH_old98}. In this case, the query time is bounded by $\mathcal{O}(3L)=\mathcal{O}(n^\rho \cdot \ln\frac{1}{\delta})$=$\mathcal{O}(n^\rho)$. However, if all the collided strings are checked, the query time can be as high as $\theta(n)$ in the worst case. But, for many real data sets, the algorithm results in sublinear query time with proper selection of the parameter values \cite{LSH_new}.

\bibliography{Reference}

%\noindent LaTeX formats citations and references automatically using the bibliography records in your .bib file, which you can edit via the project menu. Use the cite command for an inline citation, e.g.  \cite{Figueredo:2009dg}.

%\section*{Acknowledgements (not compulsory)}

%Acknowledgements should be brief, and should not include thanks to anonymous referees and editors, or effusive comments. Grant or contribution numbers may be acknowledged.

\section*{Author contributions statement}

A.C. conceived the idea, carried out the work, wrote the main text, prepared the Figures
and the software tool for conLSH. S.B. planned the work, provided laboratory facilities and wrote the manuscript. Both authors reviewed the manuscript.
\section*{Additional information}

 \textbf{Competing financial interests:}
The authors declare no competing financial interests.

\section*{Acknowledgment}
SB acknowledges JC Bose Fellowship Grant No. SB/SJ/JCB-033/201.6 dated 01/02/2017 of DST, Govt. of India and DBT funded project Multi-dimensional Research to Enable Systems Medicine: Acceleration Using a Cluster Approach for partially supporting the study. A part of this work was done when SB was visiting ICTP, Trieste, Italy.

%\begin{figure}[ht]
%\centering
%\includegraphics[width=\linewidth]{conLSH.eps}
%\caption{conLSH Vs LSH, Supplementary Figure (placed here temporarily)}
%\label{conLSH}

% \begin{table}[ht]
% \centering
% \begin{tabular}{|l|l|l|}
% \hline
% Condition & n & p \\
% \hline
% A & 5 & 0.1 \\
% \hline
% B & 10 & 0.01 \\
% \hline
% \end{tabular}
% \caption{\label{tab:example}Legend (350 words max). Example legend text.}
% \end{table}

% Figures and tables can be referenced in LaTeX using the ref command, e.g. Figure \ref{fig:stream} and Table \ref{tab:example}.

\end{document}